\documentclass[aps,prl,twocolumn,10pt,superscriptaddress,showpacs]{revtex4-2}
\usepackage{xcolor}
\usepackage{enumerate}
\usepackage{graphicx}% Include figure files
\usepackage{dcolumn}% Align table columns on decimal point
\usepackage{bm}% bold math
\usepackage[version=3]{mhchem} % Formula subscripts using \ce{}
\usepackage{amsmath,amssymb}
\DeclareMathOperator{\arccot}{arccot}
\usepackage{esvect}
\usepackage{array}
\usepackage{pifont}
\usepackage{natbib}
\usepackage{mathtools}
\usepackage{slashed}
\usepackage{nicefrac}

\usepackage{color}

\usepackage{hyperref}
\hypersetup{
  colorlinks = true,
  urlcolor   = blue,
  linkcolor  = blue,
  citecolor  = blue
}

\begin{document}

\title{Hilbert-P\'{o}lya conjecture via critical pseudo-magnetic degrees of freedom}

\author{Godwill Mbiti Kanyolo}
%\email{gmkanyolo@gmail.com; gmkanyolo@mail.uec.jp; gm.kanyolo@aist.go.jp}
\email{gmkanyolo@gmail.com; gm.kanyolo@aist.go.jp}
\affiliation{Research Institute of Electrochemical Energy (RIECEN), National Institute of Advanced Industrial Science and Technology (AIST), 1-8-31 Midorigaoka, Ikeda, Osaka 563-8577, Japan}
%\affiliation{The University of Electro-Communications, Department of Engineering Science,\\ 
%1-5-1 Chofugaoka, Chofu, Tokyo 182-8585, Japan}

\author{Titus Masese}
\email{titusmasese@gmail.com; titus.masese@aist.go.jp}
\affiliation{Research Institute of Electrochemical Energy (RIECEN), National Institute of Advanced Industrial Science and Technology (AIST), 1-8-31 Midorigaoka, Ikeda, Osaka 563-8577, Japan}
%\affiliation{AIST-Kyoto University Chemical Energy Materials Open Innovation Laboratory (ChEM-OIL), Yoshidahonmachi, Sakyo-ku, Kyoto-shi 606-8501, Japan}

\begin{abstract}
Motivated by a recent pseudo-spin model for monolayer-bilayer phase transitions in silver-based honeycomb layered materials, we propose that the critical pseudo-magnetic fields in such systems correspond to both the infinite-channel Feshbach resonance widths of a (Fermi-Dirac/Bose-Einstein/\textit{etc.}) condensate in $2$ dimensions, and equivalently to the Lee-Yang zeros of the Ising model of two pseudo-spins with a partition function corresponding to a class of functions that must include the Riemann
Xi function. Identifying the quantum-mechanical operator that yields the \textit{discontinuous/random/topological} spectrum of the critical pseudo-magnetic fields in such systems offers a tenable realisation of the Hilbert-P\'{o}lya conjecture.
\end{abstract}

\maketitle
%\tableofcontents

\textit{\textbf{Introduction.\textemdash}} Theories of critical phenomena\cite{stanley1999scaling} in unbounded $D \leq 2$ dimensional (1D/2D) systems at finite temperature such as Ising, Heisenberg, Kitaev and Berezinskii-Kosterlitz-Thouless models must be characterised by a \textit{discontinuous/random/topological} spectrum of any spontaneously broken symmetry in the phase transition since the Mermin-Wagner-Hohenberg/Coleman theorem otherwise precludes \textit{continuous} spontaneous symmetry breaking in $D < 3$.\cite{coleman1973there, hohenberg1967existence, mermin1966absence} For instance, it is known that this theorem, which precludes formation of Fermi-Dirac/Bose-Einstein/\textit{etc.} condensates in 2D\cite{schumayer2011colloquium}, manifests as the failure of analytic continuation of the Riemann zeta function\cite{broughan2017equivalents, conrey2015riemann, edwards2001riemann, keiper1992power, titchmarsh1986theory}, 
\begin{subequations}\label{zeta_eq}
    \begin{align}
        \zeta(s) = \sum_{n \in \mathbb{Z}_{> 0}} \frac{1}{n^s} = \prod_{p \in primes} \frac{1}{(1 - p^{-s})},\,\,\Re\{s\} > 1,\\
        \zeta(s) = 2^s\pi^{s - 1}\sin \left (\frac{\pi s}{2}\right)\Gamma(1 - s)\zeta(1 - s),
    \end{align}
\end{subequations}
at $s = 1$, where the boson number density/order parameter scaling as\cite{schumayer2011colloquium},
\begin{align}\label{O_D_eq}
    \mathcal{O}_D \sim \sum_{n \in \mathbb{Z}_{> 0}} \frac{1}{n^{D/2}},
\end{align}
diverges for $D < 3$, with $D \in \mathbb{Z}_{\geq 1}$ a positive integer. Here, $\Gamma(s)$ is the Gamma function. Nonetheless, since the 1D order parameter $\mathcal{O}_{D = 1} \sim \zeta(1/2) < 0$ can be regularised via zeta function regularisation\cite{hawking1977zeta} to be finite \textit{albeit} negative\cite{matsuoka1979values}, we might suspect that $\mathcal{O}_{D = 1}$ plays a role in characterising intriguing \textit{discontinuous/random/topological} critical phenomena. 

Herein, starting with the order parameter for a Fermi-Dirac/Bose-Einstein/\textit{etc.} condensate obeying the Gross-Pitaevskii equation\cite{schumayer2011colloquium} in three dimensions (3D) with a potential energy term restricted to 2D, we consider a separation of variables such that the 3D order parameter reduces to the Ginzburg-Landau equation\cite{tinkham2004introduction} in 2D, with a scattering length that depends on the 1D order parameter $\phi(z, t = 0)$ and obeys the backwards heat equation\cite{rodgers2020bruijn}, 
\begin{align}\label{backwards_heat_eq}
    \frac{\partial}{\partial t} \phi(z, t) = -\frac{\partial^2}{\partial z^2}\phi(z, t).
\end{align}
Here, $z = Z/\ell \coloneqq ih$ is the re-scaled coordinate in the $Z$ direction, also appearing in the Laplacian of the 3D Gross-Pitaevskii equation (normal to the other spatial coordinates constituting the 2D plane) after the separation of variables and $t = -icT/\ell$ is the time coordinate, $T$ after Wick-rotation\cite{zee2010quantum}, with all coordinates re-scaled with the reduced Compton wavelength, $\ell$ and speed of light $c$ in the sample, in order to work with dimensionless units. Taking the initial condition to correspond to the Riemann Xi function\cite{conrey2015riemann, edwards2001riemann, rodgers2020bruijn, csordas1994lehmer}, $\xi(s = 1/2 + iz/2) = \Xi(z/2) = \Xi(0)\phi(z, t = 0)$ with an appropriate normalisation, $\Xi(0) = \xi(1/2)$, we capture a scenario where the essential zeros $z_g/2$ of the Riemann zeta function\cite{titchmarsh1986theory, odlyzko1987distribution} satisfying $\xi(1/2 + iz_g/2) = \Xi(z_g/2) = 0$ form a \textit{discontinuous/random/topological} spectrum of critical pseudo-magnetic fields, $h_g = -iz_g$ (with $z_g \in \mathbb{R}$) in a phase transition characterised by the 1D Ising model (with two pseudo-spins subject to a periodic boundary condition), without any violation of the Mermin-Wagner-Hohenberg/Coleman theorem for the system in $D = 1,2$ ($D < 3$) dimensions.\cite{coleman1973there, hohenberg1967existence, mermin1966absence} 

Moreover, the Hadamard representation\cite{edwards2001riemann} of $\xi(s)$ with the product performed over pairs of zeros, $\varrho_g = 1/2 + iz_g/2$ and $1 - \varrho_g = 1/2 - iz_g/2$ allows the crucial link to the physics of Feshbach resonance\cite{chin2010feshbach, ketterle2008making}, 
\begin{subequations}\label{Feshbach_eq}
\begin{align}
    \phi\left (\frac{1}{\sqrt{B/B_0 - 1}}, t = 0\right ) = \lim_{N \rightarrow +\infty}\prod_{g = 1}^{N/2}\left (1 - \frac{W_g}{B - B_0}\right ), 
\end{align}
where $B \in \mathbb{R} \rightarrow \mathbb{C}$ plays the role of the externally applied magnetic field, $0 < B_0 \in \mathbb{R}$ is the magnetic field strength where Feshbach resonance occurs and $W_g = B_0/z_g^2$ correspond to $N/2$-number of resonance widths as $N \rightarrow +\infty$. The 2D scattering length vanishes for critical magnetic fields, $B_g = B_0 + W_g$ corresponding to, 
\begin{align}
    B_g = B_0\left (1 + \frac{1}{z_g^2}\right),
\end{align}
\end{subequations}
with the definition, $z_{g = 0} \rightarrow \pm\infty$. The specific condensed matter physics considerations necessary to realise such a system experimentally in silver-based honeycomb bilayered materials via a structural (bilayer-to-monolayer/3D-to-2D) phase transition induced by critical magnetic/pseudo-magnetic degree of freedom has been hinted and explored elsewhere.\cite{kanyolo2023pseudo} 

We observe that, identifying a Hermitian quantum-mechanical operator that yields the \textit{discontinuous/random/topological} spectrum of the critical magnetic fields, $B_g/B_0$ and/or the critical pseudo-magnetic fields, $z_g$ in such systems with $\zeta(1/2 \pm iz_g/2) = 0$ offers a tenable realisation of the Hilbert-P\'{o}lya conjecture, which asserts that, $z_g$ correspond to the eigenvalues of an $N\times N$ Hermitian matrix, $\gamma_N^{\dagger} = \gamma_N$ (or a Hermitian elliptic operator, $\gamma^{\dagger} = \gamma$) acting on an $N$-dimensional Hilbert space in the large $N \rightarrow +\infty$ limit of an unidentified spectral theory such as standard quantum mechanics (or quantum field theory).\cite{schumayer2011colloquium} 

Finally, we discuss whether (or not) our methods and observations are sufficient to ultimately resolve this outstanding conjecture to the affirmative. Despite the theoretical physics formalism employed throughout, we anticipate no significant obstacles to number theorists successfully delving into the details showcased in the present work.

\textit{\textbf{Theoretical Model.\textemdash}} To begin, consider the well-known Gross-Pitaevskii equation in 3D\cite{schumayer2011colloquium},
\begin{multline}\label{Gross_P_eq}
    i\hbar\frac{\partial \varphi_{\rm 3D}(\vec{X}, T)}{\partial T} = -\frac{\hbar^2}{2m}\nabla_{\vec{X}}^2\varphi_{\rm 3D}(\vec{X}, T)\\
    - E(\vec{R})\varphi_{\rm 3D}(\vec{X}, T) \pm \frac{4\pi\hbar^2a_{\rm 3D}}{m}|\varphi_{\rm 3D}(\vec{X}, T)|^2\varphi_{\rm 3D}(\vec{X}, T),
\end{multline}
describing a superfluid (Fermi-Dirac/Bose-Einstein/\textit{etc.} condensate) of bosons including paired fermions (Cooper pairs) with order parameter $\varphi_{\rm 3D}(\vec{X}, T)$ interacting via a potential $E(\vec{R})$ varying only in the $X$ and $Y$ directions, where the vectors $\vec{R} = (X, Y)$ and $\vec{X} = (X, Y, Z)$ live in two and three dimensions (2D and 3D) respectively, $T$ is coordinate time, $\hbar$ is the Planck constant, $i = \sqrt{-1}$, $m = M/2$ is the reduced mass of the paired fermions (each fermion with mass, $M$), the Laplacian is given by $\nabla_{\vec{X}}^2 = \partial^2/\partial X^2 + \partial^2/\partial Y^2 + \partial^2/\partial Z^2$ and $0 < a_{\rm 3D} \in \mathbb{R}$ is the 3D/background scattering length with $+/-$ for a Fermi-Dirac/Bose-Einstein/\textit{etc.} condensate.\cite{chin2010feshbach, ketterle2008making} 

Since the (2D) potential, $E(\vec{R})$ only varies over the $X-Y$ plane, we can perform a separation of variables,
\begin{subequations}\label{separation_eq}
\begin{align}
    \varphi_{\rm 3D}(\vec{X}, T) = \varphi_{\rm 2D} (\vec{R})\phi(Z, T),
\end{align}
which yields a pair of quasi-decoupled equations,
\begin{multline}\label{Gross_P_eq2}
   E(\vec{R})\varphi_{\rm 2D} (\vec{R}) = -\frac{\hbar^2}{2m}\nabla_{\vec{R}}^2\varphi_{\rm 2D} (\vec{R})\\
    \pm \frac{4\pi\hbar^2a_{\rm 2D}\overline{a_{\rm 2D}}}{m}|\varphi_{\rm 2D} (\vec{R})|^2\varphi_{\rm 2D} (\vec{R}),
\end{multline}
and,
\begin{align}\label{H_PDE_eq}
    i\hbar\frac{\partial}{\partial T}\phi(Z, T) = -\frac{\hbar^2}{2m}\frac{\partial^2}{\partial Z^2}\phi(Z, T),
\end{align}
\end{subequations}
corresponding to a Ginzburg-Landau equation\cite{tinkham2004introduction} in 2D and a free particle Schr\"{o}dinger equation in 1D respectively, where $\nabla_{\vec{R}}^2 = \partial^2/\partial X^2 + \partial^2/\partial Y^2$ is the 2D Laplacian and,
\begin{align}\label{2D_scat_length_eq}
   a_{\rm 2D} = \phi(Z, T)\sqrt{a_{\rm 3D}},
\end{align}
and its complex-conjugate, $\overline{a_{\rm 2D}}$ play the role of a 2D scattering length varying in time, $T$ along the $Z$ coordinate. Evidently, the decomposition of eq. (\ref{Gross_P_eq}) into eq. (\ref{Gross_P_eq2}) and eq. (\ref{H_PDE_eq}) by the separation of variables given in eq. (\ref{separation_eq}) is not unique, but rather, it is conveniently chosen in order for $|\phi(Z, T)|^2$ to be treated as the 1D number density re-scaling the 3D/background scattering length, $a_{\rm 3D}$. Indeed, one can assume the separation of variables in eq. (\ref{separation_eq}) for the condensate in eq. (\ref{Gross_P_eq}) is relevant for 1D/2D quantum dynamics such as in linear chains (\textit{e.g.} a Peierls' distorted material such as polyacetylene\cite{zee2010quantum, gontier2023phase, laughlin1999nobel} or charge solitons in a linear array of small Josephson junctions\cite{kanyolo2021renormalization, kanyolo2020rescaling, kanyolo2020cooper}) or a honeycomb layered material (\textit{e.g.} graphene\cite{zubkov2015emergent, semenoff1984condensed} and silver-based bilayered structures\cite{kanyolo2023pseudo, kanyolo2023honeycomb, masese2023honeycomb, kanyolo2022advances, kanyolo2023advances}). 

For instance, solitons that satisfy the $1 + 1$ D sine-Gordon equation are relativistic, with the soliton speed, $v_{\rm s}$ playing the role of the speed of light, $c$ and the plasma frequency, $\omega_{\rm p}$ introducing a mass scale, $m = \hbar\omega_{\rm p}/v_{\rm s}^2$.\cite{kanyolo2021renormalization} Meanwhile, $1 + 2$ D electron and/or cation dynamics in honeycomb systems such as graphene and honeycomb layered materials has been predicted to exhibit regions (such as Dirac points) where emergent special and/or general relativistic physics can occur, with the Fermi speed, $v_{\rm F}$ playing the role of the speed of light, $c$.\cite{zubkov2015emergent, kanyolo2020idealised, kanyolo2022cationic} Consequently, in addition to applying a Wick rotation (corresponding to considering the system to be at finite temperature), henceforth, we re-scale the quantities in eq. (\ref{H_PDE_eq}) as,
\begin{subequations}\label{dimensionless_eq}
\begin{align}
    \pm icT/\ell \rightarrow t,\,\,
    z = Z/\ell,\\
    a_{\rm 2D} = \sqrt{a_{\rm 3D}}\,\phi(Z, T) \rightarrow \phi_{\pm}(z, t),
\end{align}
\end{subequations}
where $\ell = \hbar/mc$ is the reduced Compton wavelength, thus obtaining the forward/backward ($+/-$) heat equation,
\begin{align}\label{heat_eq}
    \frac{\partial}{\partial t}\phi_{\pm}(z, t) = \pm \frac{\partial^2}{\partial z^2}\phi_{\pm}(z, t).
\end{align}
Crucially, despite $T = \overline{T}$ being a real parameter, $\overline{t} = t$ is also a real parameter in spite of the validity of eq. (\ref{dimensionless_eq}), since the Wick rotation only represents a transformation of the underlying metric signature of the theory from Minkowski (zero temperature) to Euclidean (finite temperature).\cite{zee2010quantum} 

We write the two general solutions for eq. (\ref{heat_eq}) as a Fourier transform, 
\begin{multline}\label{H_gen_sol_eq}
    \phi_{\pm}(z, t) = \int_{-\infty}^{+\infty}F^{-1}[\phi](u)\exp(\mp u^2t)\exp(iuz)du,
\end{multline}
where we have introduced the function $F^{-1}[\phi](u)$ of the real variable, $u \in \mathbb{R}$ and $F^{-1}$ signifies the inverse Fourier transform. As expected, the backward and forward solutions are related by time-reversal, $t \rightarrow -t$,
\begin{align}\label{time_reversal_eq}
    \phi_{\mp}(z, t) = \phi_{\pm}(z, -t),
\end{align}
with the caveat that, the backward heat equation may not be a well-posed Cauchy problem.\cite{zhang2020note, miranker1961well} Nonetheless, since we also expect the functions $\phi_+(z, t)$ and $\phi_-(z, t)$ to coincide at $t = 0$, we define $\phi_-(z, 0) = \phi_+(z, 0) \coloneqq \phi(z)$ and identify $F^{-1}[\phi_{\pm}](u, t = 0) \coloneqq F^{-1}[\phi](u)$ with the inverse Fourier transform,
\begin{align}\label{psi_Fourier_eq}
    F^{-1}[\phi_{\pm}](u, t) = \frac{1}{2\pi}\int_{-\infty}^{+\infty} \phi_{\pm}(z, t)\exp(-iuz)dz,
\end{align}
of $\phi(z)$ at $t = 0$, to obtain,
\begin{align}
    F^{-1}[\phi_{\pm}](u, t) = F^{-1}[\phi](u)\exp(\mp u^2t).
\end{align}
Finally, since the coordinate $z$ is real-valued, one can make use of the heat kernel in $D$-dimensional Euclidean/flat space ($\vec{z} \in \mathbb{R}^D$),
\begin{align}\label{heat_kernel_eq}
    K_D(\vec{z}, \vec{z'}, t) = \frac{\exp(-||\vec{z} - \vec{z'}||^2/4t)}{(4\pi t)^{D/2}},
\end{align}
to write the general solution, $\phi_+(z, t)$ for the forward heat equation in eq. (\ref{H_gen_sol_eq}) as, 
\begin{align}
    \phi_+(z, t) = \int_{-\infty}^{+\infty} dz'\,K_{D = 1}(z, z', t)\phi_+(z', 0).  
\end{align}

%\section{Results}

%\textit{\textbf{Two-pseudo-spin Ising model.\textemdash}} 
To make further progress, we shall exclusively work with the backward heat equation.\cite{rodgers2020bruijn, csordas1994lehmer} Thus, for convenience of notation, we henceforth drop the $-$ sign by taking $\phi_-(z, t) \equiv \phi(z, t)$. %\textit{\textbf{Feshbach resonance.\textemdash}} 
Proceeding, the fact that the 2D scattering length, $a_{\rm 2D}$ appearing in eq. (\ref{2D_scat_length_eq}) depends on $\phi(Z, T)$ (and hence $\phi(z, t)$ after coordinate re-scaling) is rather peculiar. Nonetheless, we can immediately establish that it vanishes ($a_{\rm 2D} = 0$) at the critical point $\phi(Z, T) = 0$ of the 3D Gross-Pitaevskii theory (\textit{i.e.} $\varphi_{\rm 3D}(\vec{X}, T) = 0$ for $1/\varphi_{\rm 2D}(\vec{R}) \neq 0$ in eq. (\ref{separation_eq})) associated with the specific values of the $Z$ coordinate (the zeros of $\phi(z)$), implying $a_{\rm 2D}$ is a candidate for the order parameter associated with a previously predicted monolayer-bilayer phase transition in silver-bilayered structures\cite{kanyolo2023pseudo}, provided the coordinate, $z \propto h$ scales as a (pseudo-)magnetic field, $h$ in the $Z$ direction. Specifically, such a re-scaling of the background (in our case, 3D) scattering length, $a_{\rm 3D}$ is reminiscent of Feshbach resonance, where the zeros $\phi(z_g) = 0$ play the role of multi-channel Feshbach resonance widths.\cite{chin2010feshbach, ketterle2008making} 

Now, consider the Hamiltonian for the Ising model with two pseudo-spins\cite{kanyolo2022cationic},
\begin{subequations}\label{Ising_eq}
\begin{align}
    H_t(u,h) = -J_t(u)\mathbf{s}_1\mathbf{s}_2 - h(\mathbf{s}_1 + \mathbf{s}_2),\\
    J_t(u) = \frac{1}{u}\ln (F^{-1}[\phi](u, t)),
\end{align}
\end{subequations}
where $\mathbf{s}_1, \mathbf{s}_2$ are single-coordinate pseudo-spin Hermitian quantum operators, $J_t(u)$ is the Heisenberg term, $t$ is taken to be a parameter hence relegated to a subscript, and $h$ is the pseudo-magnetic field. Treating $u$ as the inverse temperature, we calculate the partition function for eq. (\ref{Ising_eq}) by mapping the problem to the exactly soluble 1D Ising model\cite{grosso2013solid},
\begin{align}
    H_t^{\rm Ising}(u, h, n) = -J_t(u)\sum_{\langle i,j \rangle}\mathbf{s}_i\mathbf{s}_j - h\sum_{m = 1}^{n + 1}\mathbf{s}_i
\end{align}
of $n$ interacting spins ($\langle i,j \rangle$ is for sum over nearest neighbour) under periodic boundary conditions, $\mathbf{s}_{n + i} = \mathbf{s}_i$ and set $n = 1$ (effectively becoming a theory of a self-interacting single pseudo-spin). Thus, we find,
\begin{subequations}\label{Trace_eq}
\begin{multline}
    \mathcal{Z}_t(u, h) = \mathcal{Z}_t^{\rm Ising}(u, h, n = 1) = {\rm Tr}[M^{n = 1}]\\
    = {\rm tr}\left  [\exp(-uH_t(u, h)) \right ]
    = 2F^{-1}[\phi](u, t)\cosh (hu).
\end{multline}
where the trace, tr is taken over the two pseudo-spins, whereas the other trace, Tr is performed over the transfer matrix of the 1D Ising model\cite{grosso2013solid}, 
\begin{align}
    M = \begin{pmatrix}
\exp(u\left[J_t(u) + h\right ]) & \exp(-uJ_t(u))\\ 
\exp(-uJ_t(u)) & \exp(u\left[J_t(u) - h\right ]) 
\end{pmatrix}.
\end{align}
\end{subequations}
Since $\exp(u^2t) = \exp((-u)^2t)$, provided $F^{-1}[\phi](u, t) = F^{-1}[\phi](-u, t)$ is an even function of $u$ \textit{i.e.},
\begin{align}\label{even_eq}
    F^{-1}[\phi](u) = F^{-1}[\phi](-u),
\end{align}
the integral, 
\begin{multline}\label{Lee_Yang_eq}
    \phi(ih, t) \coloneqq \lim_{u' \rightarrow +\infty} \int_0^{u'}\mathcal{Z}_t(u, h)\,du\\
    = \lim_{u' \rightarrow +\infty} \frac{1}{2}\int_{-u'}^{u'}\mathcal{Z}_t(u, h)du = \frac{1}{2}\int_{-\infty}^{+\infty}\mathcal{Z}_t(u, h)du, 
\end{multline}
is equivalent to the general solution for the backward heat equation given in eq. (\ref{H_gen_sol_eq}) when $h = \pm iz$ (herein, we choose $h = -iz$), where $u$ is dimensionless and the limit $u' \rightarrow +\infty$ implies we are considering the theory at zero absolute temperature (corresponding to the ground state of the theory).

\textit{\textbf{Results.\textemdash}} Naturally, the physics described will depend on the choice of initial condition on $F^{-1}[\phi](u, t)$, specifically, its functional form, $F^{-1}[\phi](u, t = 0) = F^{-1}[\phi](u)$ at the time-coordinate origin, $t = 0$. This translates into studying the class of functions satisfying the heat equations given in eq. (\ref{heat_eq}), with each function studied corresponding to a different physical system. 

Assuming the function, $\phi_-(z, t = 0) = \phi_+(z, t = 0) = \phi(z) \in C^{\infty}(\mathbb{R})$ at $t = 0$ is a smooth and continuous analytic function of $z \in \mathbb{R}$ on the real line with real coefficients, we can extend it to a smooth and continuous holomorphic function on the closed upper-half plane, $\mathbb{H}\cup\partial\mathbb{H} = \{\,z'\,| \,\Im(z')\,\geq 0\}$ including the real line, $\mathbb{R} \equiv \partial\mathbb{H}$ using the Poisson kernel method\cite{Gergun2002Poisson},
\begin{subequations}
\begin{align}
    \phi(z') = \frac{1}{\pi}\int_{\mathbb{R}} dz\,\Im\left(\frac{1}{z - z'}\right)\phi(z),
\end{align}
with $z \in \mathbb{R} \rightarrow z' \in \mathbb{H}\cup\partial\mathbb{H}$. Note that, the original, $\phi(z): \mathbb{R} \rightarrow \mathbb{R}$ can be recovered in the limit as $\Im(z') \rightarrow 0^+$ using the Sokhotski-Plemelj theorem.\cite{Shiga2013Boundary} Moreover, dropping the prime in $z'$ for brevity, the Schwarz reflection principle\cite{xiaojun1996schwarz}, 
\begin{align}
    \phi(z) = \overline{\phi(\overline{z})}, 
\end{align}
\end{subequations}
can be employed to extend it to a holomorphic function on the complex plane, $\mathbb{C}$.  

However, since the inverse Fourier transform $F^{-1}[\phi](u)$ of $\phi(z)$ is no longer well-defined for $z \neq \overline{z} \in \mathbb{C}$ (the inverse Fourier transform $F^{-1}[\phi](u)$ can only be reasonably performed for $z = \overline{z} \in \mathbb{R}$), to remember this fact, we must slightly reconsider our notation for $F^{-1}[\phi](u, t = 0) = F^{-1}[\phi](u)$ appearing in eq. (\ref{H_gen_sol_eq}). Precisely, we make the replacement, $F^{-1}[\phi](u) \rightarrow \Upsilon(u)$, where $\Upsilon(u)$ is a real-valued function satisfying,
\begin{subequations}\label{Phi_asterisk_eq}
\begin{align}
    F^{-1}[\phi](u) = \Upsilon(u),
\end{align}
whenever $z = \overline{z} \in \mathbb{R}$. In the case of interest to our present work, we make a specific normalised choice for the $\Upsilon(u)$ function,
\begin{align}
    \Upsilon(u) \coloneqq \Phi(u)/\Xi(0),
\end{align}
\end{subequations}
where $\Phi(u) \geq 0$ is the super-exponentially decaying function\cite{rodgers2020bruijn},
\begin{subequations}\label{even_eq2}
\begin{align}
     \Phi(u) = \tau f(\tau) = \tau^{-1}f(1/\tau) = \Phi(-u),\\
     f(\tau) = 4\sum_{n = 1}^{+\infty}\left (2\pi^2n^4\tau^8 - 3\pi n^2\tau^4\right )\exp(-\pi n^2\tau^4),\\
     \tau (u) = \exp(-u),
\end{align}
\end{subequations}
with $\Xi(0) = \overline{\Xi(0)}$ a real-valued proportionality constant (to be subsequently computed) and $\tau$ a re-scaled $u$-coordinate. Evidently, the super-exponentially decaying function, $\Phi(u)$ in eq. (\ref{even_eq2}) makes the crucial connection to the Riemann zeta function.\cite{rodgers2020bruijn}  

Thus, taking the initial conditions given in eq. (\ref{Phi_asterisk_eq}) and eq. (\ref{even_eq2}), it is clear that, $\phi(z, 0) \coloneqq \phi(z)$ is proportional to Riemann $\xi$ function\cite{edwards2001riemann, conrey2015riemann},  
\begin{multline}\label{Riemman_xi_eq}
    \xi\left (s\right ) = (s - 1)\pi^{-s/2}\Gamma (s/2 + 1)\zeta(s)\\
    = \frac{1}{2}\prod_{\varrho}\left ( 1 - \frac{s}{\varrho}\right )\left ( 1 - \frac{s}{1 - \varrho}\right )\\
    = \lim_{u' \rightarrow +\infty}\int_{1/\tau(u')}^{\tau(u')}\frac{d\tau}{\tau}\tau^{2s}f(\tau) = \Xi(0) \phi(z) \coloneqq \Xi(z/2),
\end{multline}
where $s = 1/2 + iz/2$, 
\begin{align}\label{Gamma_funct_eq}
    \Gamma (w) = \int_0^{+\infty} \frac{dt}{t}t^w\exp(-t),
\end{align}
is the Gamma function ($w = s/2 + 1$), $\Xi(z/2) \coloneqq \xi\left (1/2 + iz/2\right )$, $\tau(u') = \exp(-u')$ and $\varrho$ and $1 - \varrho$ are the essential zeros of the Riemann zeta function paired in the Hadamard product to ensure convergence\cite{edwards2001riemann}, $\pm h_g \in \mathbb{C}$ are the zeros of the entire function,
\begin{multline}\label{xi_entire_eq}
    \Xi(ih/2) = \lim_{u' \rightarrow +\infty} \int_{-u'}^{u'} du\,\Phi(u)\exp(-uh)\\
    = \lim_{N \rightarrow +\infty} \sum_{n = 0}^N \left [ \frac{1}{n!}\int_{-\infty}^{+\infty}du\,\Phi(u)u^n\right ](-1)^nh^n\\
    \coloneqq \lim_{N \rightarrow +\infty} \sum_{n = 0}^Ng_{N - n}(-1)^nh^n\\
    \coloneqq \lim_{N \rightarrow +\infty} g_N\sum_{n = 0}^Nb_n(-1)^nh^n\\
    = \lim_{N \rightarrow +\infty} g_N\phi_N(ih) = g_{\infty}\phi(ih),
\end{multline}
with $h = -iz$ and, 
\begin{subequations}\label{coeff_relation_eqs}
\begin{align}
    g_{N - n} \coloneqq f_n = f_0b_n \geq 0,\\
    g_{\infty} = \Xi(0) = \xi(1/2) \coloneqq f_0 \geq 0. 
\end{align}
\end{subequations}
Since physical systems are always finite, the limit, $N \rightarrow +\infty$, while extremely convenient for the formulation of our theory, is certainly a mathematical artefact approximating a system with a large number, $N$ of degrees of freedom. Moreover, the definitions in eq. (\ref{even_eq2}) via eq. (\ref{Riemman_xi_eq}) enforce the even condition (eq. (\ref{even_eq}) for $z \in \mathbb{R}$),
\begin{align}\label{functional_eq}
    \Upsilon(u) = \Upsilon(-u),
\end{align}
equivalent to the functional equation, $\xi(s) = \xi(1 - s)$ of the Riemann zeta function, $\zeta(s)$. Particularly, since $0 \leq u \leq +\infty$ plays the role of inverse temperature in the partition function (eq. (\ref{Trace_eq})), the functional equation corroborates eq. (\ref{Lee_Yang_eq}), which analytically extends the validity of the values of the inverse temperature, $u$ to the entire real line, \textit{i.e.} $-\infty \leq u \leq +\infty$. 

Referencing eq. (\ref{coeff_relation_eqs}), since $\zeta(1/2) < 0$ is finite and negative\cite{matsuoka1979values}, and $\Gamma(1/4) > 0$ is finite and positive, we can confirm using eq. (\ref{Riemman_xi_eq}) that the coefficient $f_0$ is not only finite but also positive,
\begin{align}\label{Xi_0_eq}
    f_0 = \xi(1/2) = -\frac{1}{\pi^{1/4}}\left(\frac{1}{2}\right )^3\Gamma (1/4)\zeta(1/2) > 0.
\end{align}
Meanwhile, the entire function $\Xi(ih/2)$ in eq. (\ref{xi_entire_eq}) is the Taylor expansion around $h = 0$, whereby its real and non-negative coefficients\cite{broughan2017equivalents, griffin2022jensen} are given by,
\begin{align}\label{Xi_Coefficients}
    n!\,f_n = \left.\frac{d^n\,\Xi(ih/2)}{(-1)^n\,dh^n}\right\vert_{h = 0} = \int_{-\infty}^{+\infty}du\,u^n\,\Phi(u) \geq 0,    
\end{align}
which guarantees the odd coefficients vanish, $f_{2n + 1} = 0$ by virtue of $f(1/\tau) = \tau^2f(\tau)$ (equivalent to $\Phi(u) = \Phi(-u)$ in eq. (\ref{even_eq2})). Thus, setting $f(\tau(u)) = g(u)$ with $\tau(u) = \exp(-u)$ also given in eq. (\ref{even_eq2}), the even/non-vanishing coefficients correspond to the Gamma function averaging\cite{kanyolo2023pseudo},
\begin{subequations}\label{gamma_correlation_eq}
\begin{align}
    f_{2n} = 2\langle\,f(\tau(u))\,\rangle_{2n + 1} = 2\langle\,g(u)\,\rangle_{2n + 1} \neq 0,\\
    \langle\,g(u)\,\rangle_w \coloneqq \frac{1}{\Gamma (w)}\int_0^{+\infty}\frac{du}{u}\,u^w\exp(-u)g(u),
\end{align}
\end{subequations}
performed with the Gamma function (eq. (\ref{Gamma_funct_eq})) at odd positive integers $w = 2n + 1$. Using, $f_0 = \int_0^{+\infty} d\tau f(\tau) = g_{\infty} = \Xi(0)$ from eq. (\ref{xi_entire_eq}), we find the coefficient,
\begin{align}\label{norm_H_unity_eq}
    \phi(z = 0)  = \int_{-\infty}^{+\infty}du\,\Upsilon(u) = b_0 = 1,
\end{align}
is normalised to unity. Proceeding, since $f_n = f_0b_n$ (eq. (\ref{coeff_relation_eqs})) and $\Xi(ih/2) = \Xi(0)\phi(ih, t = 0)$ is related to the partition function $\phi(ih, t = 0)$ of the pseudo-spin model in eq. (\ref{Lee_Yang_eq}) at $t = 0$, $f_n$ can also be obtained by computing the $n$-point correlations of a pseudo-magnetic moment defined as,
\begin{align}\label{total_sigma_eq}
    \sigma \equiv \sigma(u) \coloneqq u\,\mathbf{s}_1 + u\,\mathbf{s}_2,
\end{align}
for the ground state ($u' \rightarrow +\infty$) by differentiating under the integral sign (Leibniz integral rule), 
\begin{multline}\label{coefficients_avrg_eq}
    \left\langle\,\sigma^n(u)\,\right\rangle \coloneqq \lim_{u' \rightarrow +\infty}\frac{\int_0^{u'} du\,{\rm tr}\left  [\sigma^n(u)\exp(-uH_0(u, h = 0)) \right ]}{\left (1 = \int_0^{u'}du\,{\rm tr}\left  [\exp(-uH_0(u, h = 0)) \right ]\right )}\\
    = \left. (-1)^n\frac{d^n}{dh^n}\phi(ih)\right\vert_{h = 0} = n!\,b_n \geq 0,
\end{multline}
where we have used eq. (\ref{norm_H_unity_eq}) to justify the denominator,
\begin{align}
    \lim_{u' \rightarrow +\infty} \int_0^{u'} du\,{\rm tr}\left  [\exp(-uH_0(u, 0))\right ] = \phi(0) = 1.
\end{align}
Similar to Gaussian integrals\cite{zee2010quantum}, it is clear that the $2n + 1$-point (odd) correlations vanish by virtue of $f_{2n + 1} = 0$, whereas the $2n$-point (even) correlations given by eq. (\ref{gamma_correlation_eq}) are non-vanishing and strictly positive. 

On the other hand, recalling the Weierstra{\ss} factorisation of the cosine function,
\begin{align}\label{cos_Weierstrass_eq}
    \cos\left (\pi w(u)\right )
    = \prod_{g = 1}^{+\infty}\left (1 - \frac{w^2(u)}{(g - 1/2)^2} \right ),
\end{align}
we shall take $w(u) = z\sigma(u)/\pi$. Now, defining the fugacity\cite{bena2005statistical} operator,
\begin{align}\label{fugacity_eq}
    \hat{\phi}(ih) \coloneqq \exp\left (\sigma(u)h\right ) \equiv \hat{\phi}(ih, t = 0),
\end{align}
and making use of $h = -iz$ and $\left \langle \sigma^{2n + 1}(u)\right \rangle = 0$ with $n \in \mathbb{N}$, eq. (\ref{cos_Weierstrass_eq}) can be employed to confirm,  
\begin{multline}\label{cos_sigma_h}
    \phi(ih) = \left \langle \hat{\phi}(ih) \right \rangle
    = \lim_{N \rightarrow +\infty}\left \langle\sum_{n = 0}^N \sigma^n(u)\frac{h^n}{n!}\right \rangle\\
    = \lim_{N \rightarrow +\infty}\sum_{n = 0}^N \left \langle \sigma^{2n}(u)\right \rangle (-1)^n\frac{z^{2n}}{2n!} = \left \langle \cos\left (z\sigma(u)\right ) \right \rangle\\
    = \lim_{N \rightarrow +\infty}\left\langle \prod_{g = 1}^{N/2} \left (1 - \frac{z^2}{E_g^2(u)}\right ) \right \rangle\\
    = \lim_{N \rightarrow +\infty}\left \langle \det\left (I_N + \Gamma_N^{-1}z \right ) \right \rangle\\
    = \lim_{N \rightarrow +\infty} \left \langle \exp\left (\sum_{n = 1}^{+\infty}\frac{(-1)^{n + 1}}{n} {\rm tr}[(\Gamma_N^{-1})^n]z^n\right ) \right \rangle\\
    = \lim_{N \rightarrow +\infty} \left \langle \exp\left (\sum_{n = 1}^{+\infty}\frac{(-1)^{n + 1}}{n} {\rm tr}[(\sqrt{-1}\,\Gamma_N^{-1})^n]h^n\right ) \right \rangle\\
    = \lim_{N \rightarrow +\infty} \left \langle \sum_{n = 0}^N\hat{e}_nh^n\right \rangle = \sum_{n = 0}^N\langle \hat{e}_n \rangle h^n,
\end{multline}
where,
\begin{align}\label{QHO_eq}
    E_g(u) \coloneqq \nu(u)(g - 1/2),
\end{align}
take the intriguing form of the energy of a quantised 1D harmonic oscillator with angular frequency, $\nu(u) = \pi/\sigma(u)$. Here, we have defined the $N\times N$ matrices, 
\begin{multline}\label{large_Gamma_N_eq}
\Gamma_N = 
\left(
    \begin{array}{ccccc}
    +E_1                            \\
      & -E_1  &   & \text{\large 0}\\
      &               & \ddots                \\
      & \text{\large 0} &   & +E_{N/2}\\
      &               &   &   & -E_{N/2}
    \end{array}
\right)\\
\coloneqq \frac{\Omega_N^{-1}\digamma_N^{-1}\Omega_N}{2\pi^{-1}} = 
\left(
    \begin{array}{ccccc}
    \mathcal{E}_1                            \\
      & \mathcal{E}_2  &   & \text{\large 0}\\
      &               & \ddots                \\
      & \text{\large 0} &   & \mathcal{E}_{N - 1}\\
      &               &   &   & \mathcal{E}_N
    \end{array}
\right),
\end{multline}
where $\digamma_N$ is an invertible and diagonalisable $N\times N$ Hermitian matrix, $\mathcal{E}_n$ ($1 \leq n \leq N$), $+E_g = -\mathcal{E}_{2g} = +\mathcal{E}_{2g - 1}$ and $-E_g = +\mathcal{E}_{2g} = -\mathcal{E}_{2g - 1}$ are the non-vanishing (diagonal) entries of the matrices $\Gamma_N$ and $N \rightarrow +\infty$ with $N$ strictly an even integer. Thus, the trace in eq. (\ref{cos_sigma_h}) corresponds to ${\rm tr}[(\Gamma_N^{-1})^n] = {\rm tr}[\Omega_N^{-1}(\Gamma_N^{-1})^n\Omega_N] = {\rm tr}[(\digamma_N)^n]/2\pi$ with $\Gamma_N = \Omega_N^{-1}\digamma_N^{-1}\Omega_N/2\pi$ from eq. (\ref{large_Gamma_N_eq}) with $\Gamma_N^{-1}\Gamma_N = \Gamma_N\,\Gamma_N^{-1} = \digamma_N\digamma_N^{-1} = \digamma_N^{-1}\digamma_N = I_N$, where $I_N = \Omega^{-1}\Omega = \Omega\,\Omega^{-1}$ is the $N\times N$ identity matrix and $\Omega = (\vec{\Omega}_1, \vec{\Omega}_2, \cdots, \vec{\Omega}_N)$ is the diaginalisation matrix constructed from the eigenvectors, $\vec{\Omega}_1$, $\vec{\Omega}_2$ and $\vec{\Omega}_N$ \textit{etc.} of some non-diagonalised Hermitian matrix, $\digamma_N$. Moreover, we appropriately identify the coefficients in eq. (\ref{cos_sigma_h}) given by $\sigma^n(u)/n! = \hat{e}_n$ to correspond to the elementary symmetric polynomials\cite{macdonald1998symmetric},
\begin{align}\label{e_hat_eq}
    \hat{e}_n = \lim_{N \rightarrow +\infty} \sum_{1 \leq k_1 < \cdots < k_n \leq N} \frac{1}{\mathcal{H}_{k_1}\times  \cdots \times \mathcal{H}_{k_n}},
\end{align}
where $\mathcal{H}_k \coloneqq -\sqrt{-1}\,\mathcal{E}_k$ ($1 \leq k \leq N$). Consequently, the power sums and Newton's identities\cite{macdonald1998symmetric} respectively correspond to,
\begin{subequations}\label{Newtons_identities_eq}
\begin{align}
    {\rm tr}[(\sqrt{-1}\,\Gamma_N^{-1})^n] = \sum_{k = 1}^N\left (\frac{1}{\mathcal{H}_k}\right )^n,\\
    \hat{e}_n = \lim_{N \rightarrow +\infty}\frac{1}{n}\sum_{l = 1}^n (-1)^{l - 1}\hat{e}_{n - l}{\rm tr}[(\sqrt{-1}\Gamma_N^{-1})^l],
\end{align}
\end{subequations}
with $1 \leq l \leq n \rightarrow +\infty$ and $\hat{e}_0 = 1$.

Proceeding, the fugacity operator in eq. (\ref{fugacity_eq}), with $\sigma = u(\mathbf{s}_1 + \mathbf{s}_2)$ as defined in eq. (\ref{total_sigma_eq}), depends on time, $t$ as, 
\begin{subequations}\label{fugacity_time_eq}
\begin{align}
     \hat{\phi}_{\pm}(ih, t) \coloneqq \hat{\phi}(ih)\exp(\mp u^2t),\\
     \phi_{\pm}(ih, t) = \langle \hat{\phi}_{\pm}(ih, t) \rangle,
\end{align}
\end{subequations}
in accordance with the two solutions of eq. (\ref{heat_eq}) given by eq. (\ref{H_gen_sol_eq}). Substituting, 
\begin{subequations}\label{Hermite_rescaling_eq}
\begin{align}
    u\sqrt{\pm 2t} = \tau(u, \pm t),\\
    \frac{h}{\sqrt{\pm 2t}}(\mathbf{s}_1 + \mathbf{s}_2) = \hat{\Sigma}(h, \pm t), 
\end{align}
\end{subequations}
into eq. (\ref{fugacity_time_eq}) yields the operator expansion for $\hat{\Sigma}$, 
\begin{align}\label{gen_Hermite_eq}
    \hat{\phi}_{\pm}(ih, t) = \exp\left (\tau\hat{\Sigma} - \tau^2/2\right ) = \sum_{n = 0}^{+\infty}H_n^*(\hat{\Sigma})\frac{\tau^n}{n!},
\end{align}
equivalent to the generating function of the probabilist's Hermite polynomials, $H_n^*(x)$, where $\tau \coloneqq \tau(u, \pm t)$ and $\hat{\Sigma} \coloneqq \hat{\Sigma}(h, \pm t)$.\cite{abramowitz1968handbook} Unlike the physicist's Hermite polynomials, the highest order coefficient $C_n$ of each probabilist's Hermit polynomial, $H_n^*(x) = C_nx^n - C_{n - 2}x^{n - 2} + \cdots + H_n^*(0)$ is normalised to unity ($C_n = 1, C_{n - 2} > 0$ \textit{etc.} and $H_{2k + 1}^*(0) = 0$, $H_{2k}^*(0) = (-1)^k2k!\,2^k/k!$ with $n = 2k, 2k + 1$).\cite{abramowitz1968handbook} Thus, eq. (\ref{Hermite_rescaling_eq}) guarantees the $n^{\rm th}$ power, $\sigma^n(u)$ of the total pseudo-magnetic moment, $\sigma(u)$ with coefficient $C_n = 1$ always appears in eq. (\ref{gen_Hermite_eq}) within the terms of the form,
\begin{multline}
    C_nu^n(\pm 2t)^{n/2}\times h^n(\mathbf{s}_1 + \mathbf{s}_2)^n/(\pm 2t)^{n/2}\\
    - u^n(\pm 2t)^{n/2}\times C_{n - 2}h^{n - 2}(\mathbf{s}_1 + \mathbf{s}_2)^{n - 2}/(\pm 2t)^{(n - 2)/2} + \cdots\\
    = \sigma^n(u) h^n \mp \frac{2C_{n - 2}}{h^2\sigma^2(u)}u^{n + 2}h^nt^1 + \cdots,
\end{multline}
thus reconstituting eq. (\ref{cos_sigma_h}) as $t \rightarrow 0^{\pm}$.  

Meanwhile, we make use of the Hadamard product representation of $\Xi(ih/2)$ in eq. (\ref{Riemman_xi_eq}), known to be convergent\cite{edwards2001riemann} when the product is evaluated over pairs of zeros, $\varrho_g = 1/2 + h_g/2 = 1/2 - iz_g/2$ and $1 - \varrho_g = 1/2 - h_g/2 = 1/2 + iz_g/2$ to yield, 
\begin{multline}\label{Hadamard_eq}
    \Xi(z/2 = ih/2) = \frac{1}{2}\prod_{g = 1}^{N/2}\left ( 1 - \frac{1/2 + h/2}{\varrho_g}\right )\left ( 1 - \frac{1/2 + h/2}{1 - \varrho_g}\right )\\
    = \frac{1}{2}\prod_{g = 1}^{N/2}\left (\frac{\varrho_g - 1/2 - h/2}{\varrho_g}\right )\left (\frac{1/2 - \varrho_g - h/2}{1 - \varrho_g}\right )\\
    = \lim_{N \rightarrow +\infty}\frac{1}{2}\prod_{g = 1}^{N/2} \left (\frac{h/2 - h_g/2}{1/2 + h_g/2}\right )\left (\frac{h/2 + h_g/2}{1/2 - h_g/2}\right )\\
    = \lim_{N \rightarrow +\infty}\frac{1}{2}\prod_{g = 1}^{N/2}\frac{(h - h_g)(h + h_g)}{1 - h_g^2}\\
    = \lim_{N \rightarrow +\infty}\frac{1}{2}\prod_{g = 1}^{N/2}\frac{(z_g - z)(z_g + z)}{1 + z_g^2} = \Xi(0) \phi(z),
\end{multline} 
where, by the fundamental theorem of algebra, the number of zeros $N$ is also the degree of the polynomial $\phi_N(z)$ in eq. (\ref{xi_entire_eq}) prior to performing the large $N$ limit, $\lim_{N \rightarrow +\infty} \phi_N(z) = \phi(z)$. To make the expressions in eq. (\ref{Hadamard_eq}) wieldy, one assumes the ordering scheme,
\begin{align}\label{height_ordering_eq}
    0 < \Re\{z_1\} \leq \Re\{z_2\} \leq \cdots \leq \Re\{z_{N/2}\},
\end{align}
by height $\mathbf{T} > \Re\{z_g\}$, for the essential Riemann zeros, $1 - \varrho_g = 1/2 + iz_g/2$ on the upper-half plane of the critical strip, where $N/2 \geq g \in \mathbb{Z}_{\geq 1}$ is strictly a positive integer bounded above by $N/2$.  

Moreover, by the functional equation (eq. (\ref{functional_eq})), $N$ must strictly be taken to be a positive even integer as it tends to infinity ($N \rightarrow + \infty$). Setting $\phi(z = 0) = 1$ (eq. (\ref{norm_H_unity_eq})), we obtain the zeroth coefficient, 
\begin{subequations}
\begin{align}
    \Xi(0) = g_{\infty} = \lim_{N \rightarrow +\infty} g_N > 0,\\
    g_N = \frac{1}{2}\prod_{n = 1}^{N/2}\frac{h_g^2}{h_g^2 - 1} = \frac{1}{2}\prod_{n = 1}^{N/2}\frac{z_g^2}{1 + z_g^2} \equiv \frac{1}{2}\prod_{n = 1}^{N/2}\frac{B_0}{B_g},
\end{align}
\end{subequations}
in terms of the essential zeros $z_g/2 = ih_g/2$ and $g_N$ (eq. (\ref{coeff_relation_eqs})), with the appearance of the magnetic degrees of freedom, $B_0$ and $B_g$ defined in eq. (\ref{Feshbach_eq}) heralding the physics of Feshbach resonance.\cite{chin2010feshbach, ketterle2008making} Consequently, the order parameter is given by,
\begin{multline}\label{prod_psi_eq}
    \phi(ih) = \phi(z) = \lim_{N \rightarrow +\infty}\prod_{g = 1}^{N/2}\frac{(h_g - h)(h_g + h)}{h_g^2}\\
    = \lim_{N \rightarrow +\infty}\prod_{g = 1}^{N/2} \left (1 - \frac{h^2}{h_g^2}\right ) = \lim_{N \rightarrow +\infty}\prod_{g = 1}^{N/2} \left (1 - \frac{z^2}{z_g^2}\right )\\
    = \lim_{N \rightarrow +\infty}\det\left (I_N + \gamma_N^{-1}z \right )\\
    = \lim_{N \rightarrow +\infty}  \exp\left (\sum_{n = 1}^{+\infty}\frac{(-1)^{n + 1}}{n} {\rm tr}[(\gamma_N^{-1})^n]z^n \right )\\
    = \lim_{N \rightarrow +\infty}  \exp\left (\sum_{n = 1}^{+\infty}\frac{(-1)^{n + 1}}{n} {\rm tr}[(\sqrt{-1}\gamma_N^{-1})^n]h^n \right )\\
    = \lim_{N \rightarrow +\infty}\sum_{n = 0}^Ne_n h^n \equiv \lim_{N \rightarrow +\infty}\sum_{n = 0}^N\left \langle \hat{e}_n \right \rangle h^n,
\end{multline}
which we have equated to eq. (\ref{cos_sigma_h}) in the last line. 

Moreover, without assuming Riemann hypothesis\cite{broughan2017equivalents, conrey2015riemann, edwards2001riemann, keiper1992power} (RH), the essential zeros $z_g/2 \in \mathbb{C}$ can be captured by the diagonalised $N\times N$ matrix,
\begin{subequations}\label{small_gamma_N_eq}
\begin{align}
    \gamma_N^{-1} \coloneqq \frac{1}{2\pi}\omega^{-1}\,\mathcal{F}_N\,\omega,
\end{align}
where $\omega = (\vec{\omega}_1, \vec{\omega}_2, \cdots, \vec{\omega}_N)$ is the diagonalisation matrix constructed from the eigenvectors, $\vec{\omega}_1$, $\vec{\omega}_2$ and $\vec{\omega}_N$ \textit{etc.} of some non-diagonalised trace-less matrix, $\mathcal{F}_N$, whereas,
\begin{multline}\label{gamma_N_matrix_eq}
    \gamma_N \coloneqq \left(
    \begin{array}{ccccc}
    M_1                                    \\
      & M_2             &   & \text{\large 0}\\
      &               & \ddots                \\
      & \text{\large 0} &   & M_{N - 1}           \\
      &               &   &   & M_N
    \end{array}
\right)\\
= \sqrt{-1}\left(
    \begin{array}{ccccc}
    \Lambda_1                                    \\
      & \Lambda_2             &   & \text{\large 0}\\
      &               & \ddots                \\
      & \text{\large 0} &   & \Lambda_{N - 1}           \\
      &               &   &   & \Lambda_N
    \end{array}
\right),
\end{multline}
\end{subequations}
satisfies $\gamma_N\gamma_N^{-1} = \gamma_N^{-1}\gamma_N = I_N$ with $\omega\,\omega^{-1} = \omega^{-1}\omega = I_N$, $\mathcal{F}_N^{-1}\mathcal{F}_N = \mathcal{F}_N\mathcal{F}_N^{-1} = I_N$ and $h_g = -iz_g \coloneqq -\Lambda_{2g - 1} = +\Lambda_{2g} = \sqrt{-1}\,M_{2g - 1} = -\sqrt{-1}\,M_{2g}$ the non-vanishing (diagonal) components of the matrix, $-\sqrt{-1}\,\gamma_N$. We can check that,
\begin{subequations}\label{average_e_eq}
\begin{align}
    \left \langle\,\hat{e}_n\,\right \rangle = e_n \coloneqq e_n(\pm z_1, \cdots, \pm z_{N/2}),\\
    e_n = \lim_{N \rightarrow +\infty} \sum_{1 \leq k_1 < \cdots < k_n \leq N} \frac{1}{\Lambda_{k_1}\times  \cdots \times \Lambda_{k_n}},
\end{align}
\end{subequations}
where $\left \langle\,\hat{e}_n\,\right \rangle$ correspond to the average values of the elementary symmetric polynomials given in eq. (\ref{e_hat_eq}) and the power sums, 
\begin{subequations}\label{Newton_eq}
\begin{align}
    {\rm tr}[(\sqrt{-1}\,\gamma_N^{-1})^n] = \sum_{k = 1}^N\left (\frac{1}{\Lambda_k}\right )^n,
\end{align}
satisfy Newton's identities\cite{macdonald1998symmetric},
\begin{align}\label{avrg_Newtons_identities_eq}
    \left \langle \,\hat{e}_n\, \right \rangle = \lim_{N \rightarrow +\infty} \frac{1}{n}\sum_{l = 1}^n (-1)^{l - 1}\left \langle \,\hat{e}_{n - l}\,\right \rangle {\rm tr}[(\sqrt{-1}\,\gamma_N^{-1})^l],
\end{align}
\end{subequations}
with $1 \leq l \leq n \rightarrow +\infty$ and $\left\langle\,\hat{e}_0\,\right \rangle = 1$. 

\textit{\textbf{Discussion.\textemdash}} While we already know that $\Gamma_N$ (or equivalently, $\digamma_N$) is both trace-less and Hermitian, simply equating $\Gamma_N$ to $\gamma_N$ (or equivalently $\digamma_N$ to $\mathcal{F}_N$) in eq. (\ref{small_gamma_N_eq}) spectacularly fails to prove RH as has been explicated in \textit{appendix A}, suggesting additional ingredients are required to make further progress within the framework of the pseudo-spin model. Indeed, it is often remarked that, an Euler product, alongside a Dirichlet series admitting a meromorphic continuation to the entire complex plane, a functional equation and a valid Ramanujan conjecture (\textit{i.e.} the axiomatic generalisations due to Selberg for a Dirichlet series to satisfy RH) are requisites for RH and all its generalisations to be valid.\cite{selberg1992old, kaczorowski2011structure, conrey1993selberg} 

Certainly, this sentiment is supported by the failure of RH for Davenport-and-Heilbronn-type functions that only admit a ``Dirichlet'' series expansion with a meromorphic continuation to the entire complex plane and a functional equation without a valid Euler product.\cite{balanzario2007zeros} Meanwhile, since the Riemann zeta function is the quintessential member of the Selberg class\cite{selberg1992old}, we conclude that the existence of a Dirichlet series and an Euler product require that the partition function of our pseudo-spin Ising model given in eq. (\ref{Lee_Yang_eq}) be related to the partition function of the Primon/Riemann gas, $\zeta(s)$ given in eq. (\ref{primon_free_energy_eq}), \textit{albeit} physically well-motivated in the range $1 < s \in \mathbb{R}$, corresponding to temperature values below its Hagedorn temperature, $1/s = \Theta_{\rm H} = 1$.\cite{julia1990statistical, spector1990supersymmetry} 

However, the pseudo-spin Hamiltonian in eq. (\ref{Ising_eq}) identifies $s - 1/2 = \pm h$ with a pseudo-magnetic field degree of freedom, instead of a temperature variable. Moreover, the fact that we require $h = -iz$ to be pure imaginary for consistency with eq. (\ref{H_gen_sol_eq}) raises another physical concern especially considering the Heisenberg term, $J_t(u) = \overline{J_t(u)}$ is already real-valued on the real line, $u \in \mathbb{R}$. Nonetheless, the concerns can immediately be remedied by recalling Lee-and-Yang's investigations of phase-transitions within Heisenberg ferromagnets (ferromagnets must satisfy, $J_t(u) \geq 0$).\cite{lee1952statistical, yang1952statistical, bena2005statistical, newman1974zeros} Thus, another approach for RH is to observe that, as $t \rightarrow 0$, eq. (\ref{Lee_Yang_eq}) is equivalent to eq. (\ref{prod_psi_eq}), predicting the existence of an infinitude of critical points for a phase transition corresponding to real-valued zeroes, $\overline{z}_g/2 = z_g/2 = ih_g/2$, in accordance with (Lee-Yang) circle theorem\cite{ruelle1971extension}, where $\hat{\phi}(iz)$ defined in eq. (\ref{fugacity_eq}) indeed plays the role of fugacity. However, non-trivial technicalities arise when the circle theorem is naively applied to the pseudo-spin model as presently formulated. These technicalities have been introduced and addressed in \textit{Appendix B}. 

Meanwhile, we obtain the Feshbach result given in eq. (\ref{Feshbach_eq}) by substituting,
\begin{align}\label{FQHE_eq}
    z = \frac{1}{\sqrt{B/B_0 - 1}} \coloneqq \frac{1}{\ell_{B_0}}\left(x + iy\right), 
\end{align}
into eq. (\ref{prod_psi_eq}), where we have defined new complex-valued coordinates, 
\begin{subequations}\label{suitable_coordinates_eq}
\begin{align}
    x + iy \coloneqq \frac{1}{\sqrt{1/\ell_B^2 - 1/\ell_{B_0}^2}},    
\end{align}
with $\ell_B = \sqrt{\hbar c/eB} \in \mathbb{C}$ and $\ell_{B_0} = \sqrt{\hbar c/eB_0} \in \mathbb{R}$ the magnetic lengths\cite{laughlin1999nobel}, $e$ the elementary charge of the electron, and $x \in \mathbb{R}$ and $y \in \mathbb{R}$ real-variables. Thus, the essential zeros are given by,
\begin{align}
    z_g = (x_g + iy_g)/\ell_{B_0},
\end{align}
\end{subequations}
with RH equivalent to the condition, $y_g = 0$ (\textit{i.e.} $B_0 < B_g \in \mathbb{R}$) for all essential zeros. 

Moreover, the functional equation, $\xi(s) = \xi(1 - s)$ requires both $\pm z_g \in \mathbb{C}$ and $\pm \overline{z_g} \in \mathbb{C}$ to appear as eigenvalues of $\gamma_N$ in eq. (\ref{gamma_N_matrix_eq}) independent of RH (RH is equivalent to $\pm z_g = \pm \overline{z_g} \in \mathbb{R}$). It follows that, all the power sums, ${\rm tr}[(\gamma_N^{-1})^n] \in \mathbb{R}$ with $n \in \mathbb{Z}_{\geq 1}$ are real-valued. In fact, since $\gamma_N$ is trace-less, all odd power sums, ${\rm tr}[(\gamma_N^{-1})^{2k - 1}]$ with $k \in \mathbb{Z}_{\geq 1}$ must vanish. Notice that, establishing the Hermicity of $\gamma_N$ (or equivalently $\mathcal{F}_N$) as $N \rightarrow +\infty$ by relating eq. (\ref{large_Gamma_N_eq}) to eq. (\ref{small_gamma_N_eq}) is tantamount to establishing the validity of RH. Moreover, all the power sums are non-negative if RH is valid. This observation is reminiscent of Li's criterion.\cite{li1997positivity} 

In fact, both the Li coefficients and the power sums appear separately as coefficients in the power series expansion of their respective pseudo-magnetisation functions,  
\begin{subequations}\label{magnetisation_eq}
\begin{align}
    \frac{\partial}{\partial h}\ln \phi(\sqrt{-1}\,h) = \sum_{k = 1}^{+\infty} (-1)^k{\rm tr}[(\gamma_N^{-1})^{2k}]h^{2k - 1},\\
    \frac{\partial}{\partial h_*}\ln \phi\left (\sqrt{-1}\,\frac{h_* + 1}{h_* - 1}\right ) = \sum_{n = 0}^{+\infty}\lambda_{n + 1}^*h_*^n,
\end{align}
\end{subequations}
where $1/s + 1/h_* = 1$, $s = 1/2 + h/2$ and, 
\begin{align}\label{Li_coeff_eq}
    \lambda_n^* = \lim_{N \rightarrow +\infty}\sum_{g = 1}^N\left( 1 - \left(\frac{\varrho_g - 1}{\varrho_g} \right)^n\right ),
\end{align}
are the Li coefficients.\cite{li1997positivity} Thus, we can define the angle,
\begin{subequations}\label{angle_eq}
\begin{align}
    h_* = \exp(i2\vartheta),
\end{align}
such that,
\begin{align}
    \vartheta = \arccos \sqrt{B_0/B} = \arccot z \in \mathbb{C},
\end{align}
\end{subequations}
and RH is equivalent to $\vartheta_g \in \mathbb{R}$ for all $B_g$ in eq. (\ref{Feshbach_eq}). Here, $h$ and $h_*$ are related by the M\"{o}bius transformation, 
\begin{align}
    h = \frac{h_* + 1}{h_* - 1} \equiv 
\begin{pmatrix}
\cos(\pi/4) & \sin(\pi/4)\\ 
-\sin(\pi/4) & \cos(\pi/4) 
\end{pmatrix}\cdot h_*,
\end{align}
under the $K \cong SO(2)$ matrix (with rotation angle, $-\pi/4$) belonging to the KAN/Iwasawa decomposition of $SL_2(\mathbb{R})$.\cite{kisil2012geometry, iwasawa1949some} Thus, it might be extremely meaningful if the remaining elements (\textit{i.e.} $A$ and $N$) of the decomposition also played a familiar role in the analysis of the Riemann $\xi(s)$ function. 

Finally, in order to preserve the Mermin-Wagner-Hohenberg/Coleman theorem\cite{coleman1973there, hohenberg1967existence, mermin1966absence} in the 1D phase transition described by the pseudo-spin model (eq. (\ref{Ising_eq})), the critical fields $B_g$ and $z_g$ in eq. (\ref{Feshbach_eq}) must be \textit{discontinuous/random/topological} in nature, as explicated in \textit{Appendix C}. A supplementary section (\textit{Appendix D}) has also been appended for the sake of rigour and overall completeness of the work.

\textit{\textbf{Acknowledgments}} --  The authors gratefully acknowledge Japan Society for the Promotion of Science (JSPS). 

\bibliography{Feshbach}
\bibliographystyle{apsrev4-2}

\appendix*

\section*{Appendices}

\textbf{Appendix A:} In the paper, we remarked that, simply equating $\gamma_N$ (or equivalently $\mathcal{F}_N$) to the Hermitian matrix $\Gamma_N$ (or equivalently, $\digamma_N$) as $N \rightarrow \infty$ spectacularly fails to validate RH, since such equalities can be shown to be fallacious. To specifically understand this failure, since eq. (\ref{cos_sigma_h}) and eq. (\ref{prod_psi_eq}) are equivalent, we make use of the fact that eq. (\ref{average_e_eq}) yields the inter-dependence of $g$ and $z_g$ via elementary symmetric polynomials and power sums.\cite{macdonald1998symmetric} As remarked above, since $\Gamma_N = \Gamma_N^{\dagger}$ is Hermitian, one would \textit{naively} expect that this inter-dependence shows $\gamma_N$, given in eq. (\ref{small_gamma_N_eq}), is also Hermitian. By eq. (\ref{average_e_eq}), we have,
\begin{align}
    {\rm tr}[\gamma_N] = \left \langle {\rm tr}[\Gamma_N] \right \rangle = {\rm tr}[\left \langle \Gamma_N \right \rangle] = 0, 
\end{align}
which is a valid expression. Since $\gamma_N$ and $\Gamma_N$ are both $N\times N$ diagonalised matrices, this expression is suggestive of, 
\begin{align}\label{fallacious_eq}
    \lim_{N \rightarrow +\infty} \gamma_N \overset{?}{=} \lim_{N \rightarrow +\infty} \left \langle \Gamma_N \right \rangle, 
\end{align}
which is manifestly Hermitian by virtue of $\Gamma_N = \Gamma_N^{\dagger}$. 

Unfortunately, due to the appearance of the trace (tr), eq. (\ref{fallacious_eq}) only represents a special case that need not be satisfied in our approach. Substituting eq. (\ref{total_sigma_eq}) and eq. (\ref{QHO_eq}) into eq. (\ref{fallacious_eq}), we obtain,
\begin{subequations}\label{fallacious_eq2}
\begin{align}
    z_g \overset{?}{=} \pi(g - 1/2)\int_0^{+\infty}\frac{du}{u}\,{\rm tr}\left[\frac{\exp\left (uJ_0(u)\mathbf{s}_1\mathbf{s}_2\right )}{\mathbf{s}_1 + \mathbf{s}_2}\right],
\end{align}
which can immediately be invalidated numerically by checking that the predicted ratio of any two essential zeros $z_g/2$ and $z_{g'}/2$ fallaciously must satisfy,
\begin{align}
    \frac{z_{g'}/2}{z_g/2} \overset{?}{=} \frac{g' - 1/2}{(g - 1/2)}.
\end{align}
\end{subequations}
Likewise, while ${\rm tr}[\gamma_N^{-1}] = \left \langle {\rm tr}[\Gamma_N^{-1}] \right \rangle = {\rm tr}[\left \langle \Gamma_N^{-1} \right \rangle] = 0$ is a valid expression, a similar fallacious argument arises by taking the special case,
\begin{align}\label{fallacious_eq3}
    \lim_{N \rightarrow +\infty} \gamma_N^{-1} \overset{?}{=} \lim_{N \rightarrow +\infty} \left \langle \Gamma_N^{-1} \right \rangle.
\end{align}
In fact, since $\langle \sigma(u)\rangle = 0$ vanishes identically, it is evident that $\gamma_N^{-1} \neq \left \langle \Gamma_N^{-1} \right \rangle = 0$. Thus, we unfortunately conclude that we must replace $\overset{?}{=}$ with $\neq$ in all such expressions that neglect the role of the trace (tr) in relating $\gamma_N$ to $\Gamma_N$ such as eq. (\ref{fallacious_eq}), eq. (\ref{fallacious_eq2}) and eq. (\ref{fallacious_eq3}) above. 

Nonetheless, we can be confident that all Newton's identities, $\langle \hat{e}_n\rangle = e_n$ (\textit{i.e.} $\langle \hat{e}_0\rangle = e_0 = 1$, $\langle \hat{e}_1\rangle = e_1 = 0$, $\langle \hat{e}_2\rangle = e_2$ \textit{etc.}) as defined in eq. (\ref{Newtons_identities_eq}), eq. (\ref{average_e_eq}) and eq. (\ref{Newton_eq}) are valid expressions. For instance, check that Newton's second identity ($n = 2$) guarantees that,
\begin{align}\label{Mellin_Lattice_eq}
    \sum_{g \in \mathbb{Z}_{> 0}}\frac{1}{z_g^2} = \frac{1}{\pi^2}\sum_{g \in \mathbb{Z}_{> 0}}\frac{\langle \sigma^2(u)\rangle}{(g - 1/2)^2} = \frac{\langle \sigma^2(u)\rangle}{2} = b_2 > 0, 
\end{align}
where $b_2$ is given in eq. (\ref{coefficients_avrg_eq}), the trace (tr) has been replaced by the sums over $g \in \mathbb{Z}_{> 0}$ and we have made use of the well-known result for the sum of reciprocals of squares of odd integers, $\sum_{g = 1}^{+\infty}1/(2g - 1)^2 = 3\zeta(2)/4$ with $\zeta(2) = \pi^2/6$ the solution to the Basel problem. While the expression simplifies, making use of the Basel result dissociates (from eq. (\ref{Mellin_Lattice_eq})) any connection of the essential zeros to the real-valued positive integers $g \in \mathbb{Z}_{> 0}$, exposing a sort of `conspiracy' preventing one from determining the validity of RH (via the Hermicity of $\gamma_N$) directly from the Hermicity of $\Gamma_N$. 

\textbf{Appendix B:} Applying the (Lee-Yang) circle theorem\cite{ruelle1971extension} to our pseudo-spin model is not exactly cogent, due to two non-subtle issues:
\begin{enumerate}[I.]
    \item The circle theorem can only be consistently invoked for a negative Heisenberg term (\textit{i.e.} it applies to the ferromagnetic condition, $J_{t = 0}(u) \geq 0$) in eq. (\ref{Ising_eq}), and not necessarily to the generic case, $-\infty < J_{t = 0}(u) < +\infty$; and,
    \item The circle theorem invoked for eq. (\ref{Lee_Yang_eq}) applies to the integrand, $\mathcal{Z}_{t = 0}(u, h) = \mathcal{Z}_0(u, h)$ and not necessarily to the integral, $\phi(ih) = \int_0^{+\infty}du\,\mathcal{Z}_{t = 0}(u, h)$.
\end{enumerate}
Circumventing the above issues is of prime importance since it would establish the validity of RH. 

To address issue I, instead of $\phi_{\pm}(z, t)$ in eq. (\ref{heat_eq}) with $z = ih$, we shall work with a modified order parameter, $\phi^*(ih, \pm t)$ defined by the following transformation,
\begin{multline}\label{transform_psi_eq}
    \phi(ih, -t) \rightarrow \phi^*(ih, -t)\\
    = \phi(ih, -t) + 2\pi CK_{D = 1}(ih, t),
\end{multline}
where $K_{D = 1}(z, t)$ is the heat kernel given in eq. (\ref{heat_kernel_eq}), $C > 1$ is a constant greater than $1$ and $\phi(ih, -t) = \phi_-(ih, -t) = \phi_+(ih, t)$ by the time-reversal property in eq. (\ref{time_reversal_eq}). Amongst other advantages, this procedure bypasses working directly with the backward heat equation, famously with an ill-posed Cauchy problem.\cite{miranker1961well, zhang2020note} Since the limit of $K_{D = 1}(ih, t)$ as $t \rightarrow 0^+$ tends to the Dirac delta function, $\delta (z) = \lim_{t \rightarrow 0^+} K_{D = 1}(z, t)$, $\phi^*(ih, 0) \equiv \phi^*(ih)$ is a meromorphic function with a single pole at $z = ih = 0$. Nonetheless, by $\delta(z) = 0$ for $z \neq 0$ and $\phi(0) = 1 \neq 0$, its zeros are equivalent to the zeros of the order parameter of interest, $\phi(z, 0) \equiv \phi(z)$. Using eq. (\ref{Phi_asterisk_eq}), observe that eq. (\ref{transform_psi_eq}) corresponds to the transformation, 
\begin{subequations}\label{transformation_eq}
\begin{align}
    \Phi(u) \rightarrow \Phi(u) + C\Xi(0) \coloneqq \Phi^*(u),\\
    \Upsilon(u) = \frac{\Phi(u)}{\Xi(0)} \rightarrow \frac{\Phi^*(0)}{\Xi(0)} = \Upsilon(u) + C \coloneqq \Upsilon^*(u).
\end{align}
\end{subequations}
Thus, the ferromagnetic condition ($t$ can be viewed as a parameter) for eq. (\ref{Ising_eq}) consistent with eq. (\ref{transform_psi_eq}) under the transformation in eq. (\ref{transformation_eq}) now becomes,
$J_t(u) \rightarrow \tilde{J_t}(u) = \left (\ln (C + \Upsilon(u))\right )/u + ut \geq 0$, which should be valid for all positive values of the inverse temperature variable $u \geq 0$. Rearranging, we find,
\begin{align}\label{extend_ferro_eq}
    t \geq \frac{1}{u^2}\ln (C + \Upsilon(u)), 
\end{align}
extends the inequality (for $\phi^*(ih, t)$ to satisfy the ferromagnetic condition) from $u \geq 0$ to the entire real line, $u \in \mathbb{R}$. 

Thus, since $\Upsilon^*(u) = \Upsilon^*(-u) = C + \Upsilon(u) > 1$ is an even function greater than unity (by $\Upsilon(u) \geq 0$ or $\Phi(u) = \Phi(-u) \geq 0$ and $\Xi(0) > 0$) and $\lim_{u \rightarrow \infty}\Upsilon(u) = 0$ actually vanishes at infinity (since it decays at least faster than the exponential), the extended ferromagnetic condition (eq. (\ref{extend_ferro_eq})) is valid for any $u$ on the real line, $\mathbb{R}$ if (and only if) $t \geq 0$, which encompasses the origin $t = 0$, as desired. Re-associating $\phi^*(z, t)$ with the backward heat equation, we observe that eq. (\ref{extend_ferro_eq}) identifies a ``de Bruijn-Newman constant''\cite{rodgers2020bruijn, csordas1994lehmer} at $t^*_{\rm dBN} = 0$, provided issue II is resolved. In this case, defining the de Bruijn-Newman constant as $t_{\rm dBN}$ (\textit{i.e.} the constant $t_{\rm dBN} \leq t$ such that the zeros $z_g \rightarrow z_g(t)$ of $\phi(z, t) \rightarrow \phi_t(z)$ are all real\cite{de1950roots, newman1976fourier}), for consistency we must have $t_{\rm dBN} = t^*_{\rm dBN}$. 

To address issue II, we treat $h = -iz = \partial/\partial u$ as a differential operator satisfying the position-momentum commutation relation given in eq. (\ref{uncertainty_relation_eq}). This is certainly consistent with the genesis of our model, since the heat equations in eq. (\ref{heat_eq}) are actually the 1D Schr\"{o}dinger equation before applying the Wick rotation, where $z$ is the dimensionless position coordinate in the $Z$ direction and $u$ is its canonical conjugate/dimensionless momentum. Thus, we are working in momentum/Fourier space such that, the usual partition function, $\mathcal{Z}_0(u, \partial/\partial u)$ meant to satisfy the Lee-Yang property\cite{newman2019lee} is promoted to an operator (\textit{i.e.} a function of the differential operator, $h = \partial/\partial u$) acting on the quantum-mechanical bra- and -ket states $\langle \,u\,|$ and $|\,u\,\rangle$ or $\langle \,z\,|$ and $|\,z\,\rangle$ respectively with $u \in  \mathbb{R}$, $z \in \mathbb{R} \rightarrow z \in \mathbb{C}$ satisfying, 
\begin{subequations}
\begin{align}
    \langle\, u\,|\,z\,\rangle = \exp(-iuz),\\
    1 = \lim_{u' \rightarrow +\infty}\int_{-u'}^{+u'}du\,|\,u\,\rangle\langle \,u\,|.
\end{align}
\end{subequations}
Precisely, the operator $\mathcal{Z}_0(u, \partial/\partial u)$ for the partition function of the two-pseudo-spin Ising model corresponds to a pseudo-spin trace operator annihilating the discrete set of quantum states, $|z_g\rangle$ corresponding to the computation,
\begin{multline}\label{Z_annihilator_eq}
   \langle\,z_g\,| \frac{1}{2}{\rm tr}\left[\exp(-uH_0(u, \partial/\partial u))\right]|\,z_g\,\rangle\\
   = \langle\,z_g\,|\frac{1}{2}\mathcal{Z}_0(u, \partial/\partial u)|\,z_g\,\rangle\\
    = \frac{1}{2}\int_{-\infty}^{+\infty} du\, \langle \,z_g\,|\,u\,\rangle\mathcal{Z}_0(u, \partial/\partial u)\langle \,u\,|\,z_g\,\rangle\\
    = \frac{1}{2}\int_{-\infty}^{+\infty} du\, \exp(iu\overline{z}_g)\mathcal{Z}_0(u, \partial/\partial u)\exp(-iuz_g)\\
    = \frac{1}{2}\int_{-\infty}^{+\infty} du\,\mathcal{Z}_0(u, -iz_g) = \phi(z_g) = 0,
\end{multline}
where we confirm we must have $z_g = \overline{z}_g \in \mathbb{R} \subset \mathbb{C}$ for consistency. Consequently, we have translated (Lee-Yang) circle theorem into a statement about the anti-Hermicity of the differential operator, $h = \partial/\partial u$ with purely imaginary eigenvalues acting on the bra- and -ket states, $\langle\,z_g\,|$ and $|\,z_g\,\rangle$ respectively -- a statement equivalent to Hilbert-P\'{o}lya conjecture.\cite{schumayer2011colloquium} Thus, the annihilated position eigen-states must be appropriately normalised as,
\begin{multline}
    \langle\,z_g'\,|\,z_g\,\rangle = \int_{-\infty}^{+\infty}du\, \langle\,z_g'\,|\,u\,\rangle \langle\,u\,|\,z_g\,\rangle\\
    = \int_{-\infty}^{+\infty}du\, \exp(iu(z_g' - z_g)) = 2\pi\delta(z_g' - z_g),
\end{multline}
as required by standard quantum mechanics, which conclusively addresses issue II. 

Finally, despite the theoretical physics formalism employed, we anticipate no significant obstacles to number theorists successfully making these arguments more formal and rigorous. 

\textbf{Appendix C:} Observe that the pseudo-spin partition function,
\begin{subequations}
\begin{align}
    \phi(ih) = {\rm E}_{X_{\sigma}}(\exp(\sigma(u) h)),
\end{align}
takes the form of a moment generating function\cite{curtiss1942note} (or the characteristic function\cite{bochner2005harmonic, gut2013characteristic},
\begin{align}
    \phi(z) = {\rm E}_{X_{\sigma}}(\exp(\sigma(u) iz)),
\end{align}
\end{subequations}
under Wick rotation $z = ih$) of the random variable,
\begin{align}
    X_{\sigma} = \sigma(u) = u\mathbf{s}_1 + u\mathbf{s}_2,
\end{align}
defined in eq. (\ref{total_sigma_eq}), where the expectation, $\rm E_{X_{\sigma}}(\cdot)$ is performed over the pseudo-spins $\mathbf{s}_1$ and $\mathbf{s}_2$ in addition to the continuous variable $u$ with the probability distribution, $\Upsilon(u) = \Phi(u)/\Xi(0)$ appearing in eq. (\ref{Phi_asterisk_eq}) and normalised as in eq. (\ref{norm_H_unity_eq}). Thus, $\ln {\rm E}_{X_{\sigma}}(\exp(\sigma(u) h))$ takes the form of the cumulant generating function\cite{gut2013characteristic}, and the $n^{\rm th}$ moments correspond to $n$-point correlations calculated using eq. (\ref{coefficients_avrg_eq}). For the fugacity in eq. (\ref{gen_Hermite_eq}) that depends on time, we recognise the moment generating function with a normal distribution, $\mathcal{N}(\hat{\Sigma}, 1)$ with mean $\hat{\Sigma}$ and unit variance. 

Moreover, since quantum theory requires the variable $z$ (originally the coordinate in $Z$ direction from eq. (\ref{Gross_P_eq})) to correspond to the Bohr complement\cite{luis2002complementarity} of the momentum $u$ in the $Z$ direction, we take,
\begin{subequations}\label{uncertainty_relation_eq}
\begin{align}
    z = ih = i\partial/\partial u,\\
    u = -i\partial/\partial z,
\end{align}
subject to Heisenberg uncertainty relation, 
\begin{align}
    [u, z] = -i.
\end{align}
\end{subequations}
In order to prove the randomness of the essential Riemann zeros $z = z_g$, it is sufficient (but not necessary) to conjecture that $\Upsilon(u) = \Upsilon(-u)$ is the probability measure for a random variable, $X = u \in \mathbb{R}$ with expectation,
\begin{align}\label{char_eq}
    {\rm E}_X(\exp(iuz)) = \phi(z) = \xi(1/2 + iz/2)/\xi(1/2).
\end{align}
While the Hermicity of $\phi(z): \mathbb{R} \rightarrow \mathbb{C}$ subject to quantum theory, $z = i\partial/\partial u$ (eq. (\ref{uncertainty_relation_eq})) is a desirable feature amenable to Bochner's theorem\cite{mackey1980harmonic}, unfortunately the above conjecture eludes proof since we are unable to show that the holomorphic function, $\phi(z)$ (or equivalently $\xi(s = 1/2 + iz/2)$) is absolutely integrable\cite{kollar2008powers}, as is required for the validity of Bochner's theorem.\cite{mackey1980harmonic}

The difficulty arises from the rather oscillatory behaviour of $\zeta(s = 1/2 + iz/2)$ on the critical strip, which contributes to $\xi(s = 1/2 + iz/2)$ via the functional equation given in eq. (\ref{Riemman_xi_eq}). In particular, the oscillatory behaviour is tied to the exponential term when we write the natural logarithm of $\zeta(s)$\cite{edwards2001riemann} as, 
\begin{multline}\label{primon_free_energy_eq}
    \frac{1}{s}\ln \zeta(s) = \int_2^{+\infty}\frac{dq}{q}\frac{\pi(q)}{q^s - 1} \equiv \int_1^{+\infty}\frac{dq}{q}\frac{\pi(q)}{q^s - 1}\\
    = \int_0^{+\infty} d\lambda\,\pi(q(\lambda))\frac{1}{\exp(s\lambda) - 1}\\
    = \int_0^{+\infty} d\lambda\,\pi(q(\lambda))\sum_{g = 1}^{+\infty}\exp\left (-s\lambda g\right ), 
\end{multline}
where, $q(\lambda) = \exp(\lambda)$ and $\pi(q)$ is the prime counting function with $\pi(q < 2) = 0$. From a mathematical physics point of view, eq. (\ref{primon_free_energy_eq}) is equivalent to the free energy for ``a gas of primes'' (the Primon/Riemann gas\cite{julia1990statistical, spector1990supersymmetry}) with partition function,
\begin{subequations}
\begin{align}
    \zeta(s) = \prod_{\ln \nu \in primes}\sum_{g = 1}^{+\infty}\exp(-sE_g(\nu)),
\end{align}
where the prime counting function\cite{conrey2015riemann},
\begin{align}
    \pi(q(\lambda)) = \sum_{p \leq q} 1 \sim q(\lambda)/\ln q(\lambda) = \exp(\lambda)/\lambda,
\end{align}    
\end{subequations}
for primes $p = \ln \nu \leq q = \ln \lambda$ counts to the number of quantum harmonic oscillators of energy, $E_g(\nu) = \nu (g - 1)$ ($g \in \mathbb{Z}_{\geq 1}$) below the angular frequency cut-off, $\lambda \geq \nu$.

In lieu of such impediments, it turns out that we can demonstrate (a path to proving) the randomness of the essential zeros by formulating the conjecture (assertion) in the Bohr complement\cite{luis2002complementarity}:

\textbf{\textit{Assertion}}: \textit{Define the expectation, ${\rm E}_Y(\exp(-iuz)) = \Phi(u)/\Phi(u = 0) \coloneqq P(u)$ for the variable, $Y = z \in \mathbb{R}$ corresponding to the essential Riemann zeros, $z_g/2 \in \mathbb{R}$ on the critical line and consider the (Bohr) complement $u \in \mathbb{R}$, where $\Phi(u) = \Phi(-u)$ is the super-exponentially decaying function given in eq. (\ref{even_eq2}) which satisfies $0 \leq \Phi(u) \leq \Phi(0)$. Then, ${\rm E}_Y(\exp(-iuz))$ is a characteristic function of random variables $Y$ with a normalised probability measure,
\begin{align}\label{Measure_eq}
    \mu(z) \coloneqq \phi_{\pm}(z, t = 0)\xi(1/2)/2\pi\Phi(0),
\end{align}
with $\xi(1/2) > 0$ given in eq. (\ref{Xi_0_eq}), and $\phi_{\pm}(z, t)$ defined as in eq. (\ref{heat_eq}).} 

As aforementioned, this assertion follows from the application of Bochner's theorem\cite{mackey1980harmonic}: 

\textbf{\textit{Proof}}: \textit{Check that, $P(u): \mathbb{R} \rightarrow \mathbb{C}$ is Hermitian (since $u = -i\partial/\partial z$ from eq. (\ref{uncertainty_relation_eq})), continuous at the origin ($\lim_{u \rightarrow 0^{\pm}}P(u) = P(0)$), positive definite ($P(u) = P(-u) \leq  P(0)$, for $u \in \mathbb{R}$) and normalised ($P(0) = 1$), and the inverse Fourier transform in eq. (\ref{psi_Fourier_eq}) is well-defined for $z \in \mathbb{R}$. Moreover, using the positive definite property, it is easy to see that $\int_{\delta'}^{\delta}|P(u)|du = \int_{\delta'}^{\delta}P(u)du$ is absolutely integrable on the real line, $u \in \mathbb{R}$ (\textit{i.e.} $-\infty \leq \delta' \leq u \leq \delta \leq +\infty$). Thus, $\phi_{\pm}(z, 0)\xi(1/2)/2\pi\Phi(0) = \mu(z) \in \mathbb{R}$ is the unique probability measure on $Y = z \in \mathbb{R}$,} which also happens to evolve under the 1D free particle Schr\"{o}dinger (or the heat equation after Wick rotation eq. (\ref{heat_eq})).

Thus, we ought to expect that, the fact that the statistics of the essential zeros can be modelled successfully by random matrix theory\cite{keating2000random} follows from the properties of the probability measure, $\mu(z)$ given in eq. (\ref{Measure_eq}), provided the trace-less matrix, $\gamma_N$ in eq. (\ref{gamma_N_matrix_eq}) is also taken to be Hermitian, in accordance with RH. Finally, the difficulty to employ Bochner's theorem to show the existence of a probability measure using eq. (\ref{char_eq}) can also be viewed as the uncertainty principle in action between the essential zeros and the natural logarithm of prime powers, $q^{-sg} = \exp(-s\lambda g)$ in eq. (\ref{primon_free_energy_eq}), as is also exemplified by the Weil explicit formula.\cite{hejhal1976selberg}

%\section*{Supplementary Information}

\textbf{Appendix D:} Without assuming the validity of RH ($z_g = ih_g \in \mathbb{C}$) unless stated otherwise, here are some additional relevant conjectures (assertions) and their (paths to) proofs: 

\textbf{\textit{Assertion 1}}: \textit{There exists a labelling scheme (defined up-to a choice-of-ordering such as adopted in eq. (\ref{height_ordering_eq})) that identifies (i) to (ii) as defined below: \textit{i.e.} (i) each essential zero, $\varrho_g^+ = 1/2 + iz_g/2 \in \mathbb{C}^+$ with $\Re\{z_g\} > 0$ and hence $\Im\{\varrho_g^+\} > 0$ (or equivalently each essential zero, $1 - \varrho_g^+ \coloneqq \varrho_g^- = 1/2 - iz_g/2 \in \mathbb{C}^-$ with $\Re\{z_g\} > 0$ and hence $\Im\{\varrho_g^-\} < 0$) within the critical strip $0 \leq \Re\{s\} \leq 1$ of the Riemann zeta function, $\zeta(s = \varrho_g^{\pm}) = 0$; and, (ii) each zero of the cosine function, $\cos(\pi(g - 1/2)) = 0$ for $g \in \mathbb{Z}_{\geq 1}$}.

\textbf{\textit{Proof}}: \textit{See eq. (\ref{large_Gamma_N_eq}) and eq. (\ref{gamma_N_matrix_eq}), while referencing eq. (\ref{prod_psi_eq}), eq. (\ref{cos_sigma_h}), eq. (\ref{Hadamard_eq}) and eq. (\ref{average_e_eq}). Clearly, the rank $N$ of the $N\times N$ matrix, $\Gamma_N$ is equal to the number of essential zeros, $N$ appearing as eigenvalues of $\gamma_N$}.

\textbf{\textit{Assertion 2}}: \textit{The essential zeros, $\zeta(\varrho_g^{\pm}) = 0$ with $\Re\{z_g\} > 0$ are labelled in pairs ($\varrho_g^{\pm} \in \mathbb{C}^{\pm}$) by a trivial zero $-2g$ of the Riemann zeta function, $\zeta(s)$ satisfying, $\zeta(-2g) = 0$, where $g \in \mathbb{Z}_{\geq 1}$}.

\textbf{\textit{Proof}}: \textit{This is a straightforward consequence of Assertion 1, and the fact that the trivial zeros of the Riemann zeta function are known to occur at $\zeta(-2g) = -B_{2g + 1}^+/(2g + 1)$ with $g \in \mathbb{Z}_{\geq 1}$, where $B_n^+$ is the $n^{\rm th}$ Bernoulli number, ($B_1^+ = +1/2$)}. 

\textbf{\textit{Assertion 3}}: \textit{Let $N \coloneqq N(\mathbf{T})$ in eq. (\ref{Hadamard_eq}) such that $N(\mathbf{T})/2$ is the number of essential Riemann zeros in the upper-half plane of the critical strip truncated at height $0 < \mathbf{T} \in \mathbb{R}$ (defining a finite rectangle, $\mathcal{R}$ on the upper half of the critical strip) and adopt the ordering scheme by height, $\mathbf{T}$ given in eq. (\ref{height_ordering_eq}), with $\Re\{z_{N/2}\}/2 = \mathbf{T} - \delta$ and $\delta > 0$ the infinitesimal. Then, 
\begin{multline}\label{Backlund_eq0}
    2g = \frac{1}{\pi}\arg\xi(1/2 + i(z_g^+/2 + \delta) + \epsilon_g)\\
    + \frac{1}{\pi}\arg\xi(1/2 + i(z_g^-/2 + \delta) - \epsilon_g),
\end{multline}
where $g \in \mathbb{Z}_{\geq 1}$, $\Im\{z_g^{\pm}\}/2 = \pm\epsilon_g$ (RH is equivalent to, $\epsilon_g = 0$ for all $z_g$) and, 
\begin{align}
    \xi(s) = (s - 1)\pi^{-s/2}\Gamma (s/2 + 1)\zeta(s) = \xi(1 - s),
\end{align}
is the completed Riemann zeta function as appearing in eq. (\ref{Riemman_xi_eq});} 

\textbf{\textit{Proof}}: \textit{Making use of a well-known result for the number of zeros in the critical strip (on and off the critical line)\cite{trudgian2012improved}, 
\begin{align}\label{Backlund_eq}
    \frac{1}{2}N(\mathbf{T}) = \frac{1}{\pi}\arg\xi(1/2 + i\mathbf{T}),
\end{align}
where $\arg\xi(s)$ is the argument of $\xi(s)$ (appropriately defined, for instance, as in ref. \cite{trudgian2012improved}), observe that, the ordering scheme given in eq. (\ref{height_ordering_eq}) and Assertion 1 of this paper guarantees, $N(\mathbf{T})/2 = g \in \mathbb{Z}_{\geq 1}$. Thus, $z_g^{\pm} = \Re\{z_g^{\pm}\} \pm i2\epsilon_g$ all appear in the ordered spectrum given by eq. (\ref{height_ordering_eq}) with $g \leq N/2$ and $(z_g^+/2 + \delta) - i\epsilon_g = \Re\{z_g\}/2 + \delta$. Note that, the infinitesimal, $\delta = \mathbf{T} - \Re\{z_{N/2}\}/2 > 0$ has been included in order to guarantee that all the essential zeros counted by the well-known formula rest within the rectangle, $\mathcal{R}$ without touching its boundary.} 

\textbf{\textit{Assertion 4}}: \textit{Any arbitrary choice-of-ordering can be transformed into another  (choice) via automorphisms\cite{yale1966automorphisms}, $\alpha(e_n) = e_n$, $\alpha(z_g) = z'_g$ and $\alpha(0) = 0$ acting on the degree $N$ polynomial, $\phi_N(z) = \sum_{n = 0}^N(-1)^ne_{2n}(\pm z_1, \cdots, \pm z_{N/2})z^{2n}$ on the upper-half plane of the critical strip, where $\xi(1/2 + iz/2) = \Xi(z/2) = \Xi(0)\phi_{+\infty}(z)$ is defined as in eq. (\ref{Hadamard_eq}) in the large $N$ limit ($N \rightarrow +\infty$), with the Riemann $\xi(s)$ function given in eq. (\ref{Riemman_xi_eq})};

\textbf{\textit{Proof}}: \textit{This follows from eq. (\ref{average_e_eq}), since $\alpha$ implements automorphisms of the set of elements, $E: \{z_1, \cdots, z_{N/2}$\} fixing another set of elements, $F: \{e_0, \cdots, e_N$\}, corresponding to the non-negative real-valued coefficients $e_{2k + 1} = 0$ and $0 < e_{2k} \in \mathbb{R}$ ($k \in \mathbb{Z}_{\geq 0}$) of the polynomial, $\phi(z = ih) = \sum_{n = 0}^{+\infty}e_nh^n$ appearing in eq. (\ref{xi_entire_eq}), eq. (\ref{coefficients_avrg_eq}), eq. (\ref{cos_sigma_h}) and eq. (\ref{Hadamard_eq}), \textit{i.e.} $\phi(z_g) \in F[z_g]$, with $F[z_g]$ the field extension of $F$. Thus, if one can show that this extension is finite, normal and separable over $F$, the powerful tools of Galois theory become at our disposal.\cite{milne2022galois} Note that, the group of these automorphisms ${\rm Aut}(E/F)$, includes the identity, $\alpha_0(z_g) = z'_g = z_g$}. 

\textbf{\textit{Assertion 5}}: \textit{Let $w = \Re\{w\} + \sqrt{-1}\Im\{w\} \in \mathbb{C}$ be a complex number, and define the automorphism, $\alpha^*(\Re\{w\}) = \Re\{w\}$, $\alpha^*(\Im\{w\}) = \Im\{w\}$ and $\alpha^*(+\sqrt{-1}) = -\sqrt{-1}$}. If $\alpha^* \in {\rm Aut}(E/F)$, then RH is equivalent to $\alpha^* = \alpha_0$, with $\alpha_0(z_g) = z_g$ the identity;

\textbf{\textit{Proof}}: \textit{Since $z_g/2$ is an essential zero of the Riemann zeta function, $\zeta(1/2 \pm iz_g/2) = 0$, this follows from the fact that $\alpha^*$ is complex-conjugation, whilst $\alpha_0$ is an automorphism fixing any element in $E$.\cite{yale1966automorphisms} Thus, to prove RH, one needs to unambiguously show that $\alpha^* = \alpha_0$, equivalent to proving the Hermicity of the trace-less matrix, $\gamma_N$ appearing in eq. (\ref{prod_psi_eq}), whose eigenvalues $\pm z_g$ are displayed in eq. (\ref{gamma_N_matrix_eq}).}

\textbf{\textit{Remark 1}}: It was argued in ref. \cite{francca2015transcendental} that, 
\begin{subequations}\label{Franca-LeClair_eq}
\begin{align}
    s'_g = \frac{1}{2} + i\frac{z'_g}{2},\,\,\Im\{z'_g\} = 0,\\
    \frac{1}{\pi}\arg\chi\left(s'_g\right) \overset{?}{=} g - 1/2 ,
\end{align}
\end{subequations}
is valid strictly on the critical line, $\Re\{s'_g\} = 1/2$, where,
\begin{align}
    \chi(s'_g) = \frac{2\xi(s'_g)}{s'_g(s'_g - 1)} = \frac{\Gamma(s'_g/2)\zeta(s'_g)}{\pi^{s'_g/2}} = 0,
\end{align}
is the completed Riemann zeta function. By implementing a shift by $+1/2$ as in $g - 1/2 \rightarrow g = g - 1/2 + 1/2$ in eq. (\ref{Franca-LeClair_eq}) and using the fact that $0 < s'_g(1 - s'_g) \in \mathbb{R}$, the authors obtain a result for the number of essential zeros, 
\begin{align}
    2g = \frac{2}{\pi}\arg \xi(s'_g + \delta) \coloneqq N_0(\mathbf{T}' = z'_g/2 + \delta),
\end{align}
on the critical line, which they remark is equivalent to the result for the number of essential zeros on the critical strip, $N(\mathbf{T} = \Re\{z_g\}/2 + \delta)$ given in eq. (\ref{Backlund_eq}). This led to their new criterion for RH\cite{francca2015transcendental}: \textit{RH follows from eq. (\ref{Franca-LeClair_eq}) with all the essential zeros simple if the equation admits a unique solution, $z_g/2$ for every, $g \in \mathbb{Z}_{\geq 1}$}.

\textbf{\textit{Assertion 6}}: \textit{If valid, eq. (\ref{Franca-LeClair_eq}) admits a unique solution, $z_g/2$ for every, $g \in \mathbb{Z}_{\geq 1}$, thus proving RH}. 

\textbf{\textit{Proof}}: \textit{The assertion follows directly from (proof of) Assertion 1}.

\textbf{\textit{Remark 2}}: \textit{However, it is not clear (to us) whether eq. (\ref{Franca-LeClair_eq}) is a valid expression on the critical line especially given the fact that it does not arise directly from $|\chi(1/2 + iz'_g/2)| = 0$ (as it should), but rather it arises from $\cos({\rm arg}\,\, \chi(s'_g)) \overset{?}{=} 0$.\cite{francca2015transcendental} Particularly, check that, $\chi(s') = |\chi(s')|\exp(i\,{\rm arg}\,\chi(s'))$, and hence, $\exp(i\,{\rm arg}\,\chi(s'_g)) \neq 0$. Thus, the essential zeros $s'_g$ of $\chi(s')$ within the critical strip (on and off the critical line) must come from $|\chi(s'_g)| = 0$. Hence, computing on the critical line, $z_g/2 \in \mathbb{R}$ (as assumed in ref. \cite{francca2015transcendental}), $\chi(s')$ need not share its zeros with the cosine function, $\cos({\rm arg}\,\chi(s'_g))$, where, $\chi(s') + \overline{\chi(s')} = 2|\chi(s'_g)|\cos({\rm arg}\,\chi(s'_g)) = 0$. Thus, we find eq. (\ref{Franca-LeClair_eq}) corresponding to the zeros, ${\rm arg}\,\chi(s_g) = (g - 1/2)\pi$ in $\cos({\rm arg}\,\chi(s'_g) = (g - 1/2)\pi) = 0$ is ill-motivated. Of course, the discourse above is contingent on taking, ${\rm arg}\,\chi(s') = \Im\{\ln \chi(s')\} \in \mathbb{R}$ as the definition of the argument of $\chi(s')$. As a result, by virtue of being unable to verify the validity of eq. (\ref{Franca-LeClair_eq}) by these methods, Assertion 6 falls just shy of proving RH.}

Nonetheless, we believe the numerical results in ref. \cite{francca2015transcendental} (after the shift by $1/2$) are valid, when they are consistent with the well-known formula given in eq. (\ref{Backlund_eq}). Crucially, the ingenious idea in ref. \cite{francca2015transcendental} that one could get the counting formula for the number of essential zeros $N_0(\mathbf{T})/2$ on the upper half of the \textit{critical line} by some novel method, and then compare its values and/or functional form with a well-established and/or novel formulae from number theory for the number of essential zeros $N(\mathbf{T})/2$ in the upper half rectangle, $\mathcal
R$ of the \textit{critical strip} (both numbers truncated at height, $\mathbf{T}$), for a proof or disproof of RH via $N_0(\mathbf{T}) \overset{?}{=} N(\mathbf{T})$, remains a valid enterprise.

\end{document}